\documentclass[aip,amsmath,amssymb,reprint,superscriptaddress]{revtex4-1}

\usepackage{dcolumn}
\usepackage{bm}
\usepackage[utf8]{inputenc}
\usepackage[T1]{fontenc}
\usepackage{mathptmx}
\usepackage{etoolbox}
\usepackage{chemformula}
\usepackage{graphicx}
\usepackage{natbib}
\usepackage{textgreek}
\usepackage{siunitx}
\usepackage[utf8]{inputenc}
\usepackage{amsmath}
\usepackage{xcolor}
\usepackage[colorlinks=true,linkcolor=blue,citecolor=blue,allcolors=blue]{hyperref}

\makeatletter
\def\@email#1#2{%
 \endgroup
 \patchcmd{\titleblock@produce}
  {\frontmatter@RRAPformat}
  {\frontmatter@RRAPformat{\produce@RRAP{*#1\href{mailto:#2}{#2}}}\frontmatter@RRAPformat}
  {}{}
}
\makeatother

\begin{document}

\preprint{AIP/123-QED}

\title[]{Spectroscopic imaging ellipsometry of two-dimensional TMDC heterostructures}

\newcommand{\WSI}{Walter Schottky Institut and Physics Department, Technical University of Munich, Am Coulombwall 4a, 85748 Garching, Germany}
\newcommand{\MCQST}{Munich Center of Quantum Science and Technology (MCQST), Schellingstr. 4, 80799 Munich, Germany}
\newcommand{\ECon}{Exzellenzcluster e-conversion, Technical University of Munich, Lichtenbergstr. 4a, 85748 Garching, Germany}
\newcommand{\WWU}{Institute of Physics, University of Münster, Wilhelm-Klemm-Str. 10, 48149 Münster, Germany}
\newcommand{\SoN}{Center for Soft Nanoscience (SoN), Busso-Peus-Str. 10, 48149 Münster, Germany}
\newcommand{\MIT}{Department of Materials Science and Engineering, Massachusetts Institute of Technology, Cambridge, MA 02139, USA}

\author{Florian Sigger}
    \affiliation{\WSI}
    \address{\MCQST}
    \address{\ECon}
\author{Hendrik Lambers}
    \affiliation{\WWU}
\author{Nisi Katharina}
    \affiliation{\WSI}
    \address{\MCQST}
    \address{\ECon}
\author{Julian Klein}
    \affiliation{\WSI}
    \address{\MIT}
\author{Nihit Saigal}
    \affiliation{\WWU}
\author{Alexander W. Holleitner}
    \affiliation{\WSI}
    \address{\MCQST}
    \address{\ECon}
\author{Ursula Wurstbauer}
    \address{\ECon}
    \affiliation{\WWU}
    \address{\SoN}
    \email{wurstbauer@wwu.de}

\begin{abstract}

Semiconducting two-dimensional materials and their heterostructures gained a lot of interest for applications as well as fundamental studies due to their rich optical properties. Assembly in van der Waals heterostacks can significantly alter the intrinsic optical properties as well as the wavelength-dependent absorption and emission efficiencies making a direct comparison of e.g. photoluminescence intensities difficult. Here, we determine the dielectric function for the prototypical MoSe$_2$/WSe$_2$ heterobilayer and their individual layers. Apart from a redshift of \SI{18}{\milli\electronvolt} - \SI{44}{\milli\electronvolt} of the energetically lowest interband transitions, we find that for larger energies the dielectric function can only be described by treating the van der Waals heterobilayer as a new artificial homobilayer crystal rather than a stack of individual layers. The determined dielectric functions are applied to calculate the Michelson contrast of the individual layers and the bilayer in dependence of the oxide thickness of often used Si/SiO$_2$ substrates. Our results highlight the need to consider the altered dielectric functions impacting the Michelson interference in the interpretation of intensities in optical measurements such as Raman scattering or photoluminescence.

\end{abstract}

\maketitle

The easily accessible method of cleaving bulk van der Waals (vdW) crystals to thin nano-layers \cite{Novoselov.2004} has led to the exploration of an ever expanding number of different 2D materials with a broad range of properties \cite{Novoselov.2016}. The possibility to combine these materials into vdW heterostructures has lead to new application prospects such as solar energy conversion \cite{NassiriNazif.2021}, solid state lighting \cite{Hwangbo.2022}, field-effect transistor based sensing\cite{Chen.2020} or spintronic devices \cite{Ahn.2020}. Experiments on transition metal dichalcogenide (TMDC) vdW heterostructures opened the avenue towards new fundamental research directions including Moiré excitons \cite{Seyler.2019,Tran.2019,DiHuang.2022}, single photon emitters \cite{Klein.2021,MichaelisdeVasconcellos.2022}, Mott-Hubbard physics \cite{Li.2021} or excitonic many-body states \cite{Wang.2019,Sigl.2020}. For both, applications as well as fundamental studies, a detailed knowledge of the modifications of the light-matter interaction described by the complex dielectric function in vdW heterostructures compared to the individual layers is important. The dielectric function can be determined by reflectance contrast and Kramers-Kronig constraint analysis \cite{Li.2014} as well as by  spectroscopic imaging ellipsometry (SIE) \cite{Fujiwara.2007}. Unlike reflectance contrast, SIE provides access to the dielectric functions also in spectral regions with weak or even no extinction, and the oblique angle of incidence provides access to the in-plane and out-of plane parts of the dielectric tensor \cite{Funke.2016}. Moreover, experimental determination of the two ellipsometric angles allows a Kramers-Kronig relation free determination of the dielectric functions with both high spatial resolution and a high signal sensitivity by using nulling ellipsometry \cite{Fujiwara.2007}.\\
Here, we combine these inherent advantages of spectroscopic imaging ellipsometry with large area vdW heterostructures to study the change in the dielectric function for MoSe$_2$ and WSe$_2$ when assembled in a heterobilayer. We report a redshift of the lowest energy A- and B-exciton resonances in the heterobilayer compared to the individual monolayers. Most strikingly, the higher lying spectral range of the heterobilayer can only be well described by a new material rather than a multilayer system of two stacked monolayers indicating the importance of interlayer hybridization effects. The determined dielectric functions are utilized to calculate the Michelson interference contrast for commonly used Si/SiO$_2$ substrates demonstrating its impact in the evaluation and interpretation of intensities in optical interband experiments such as photoluminescence (PL), absorbance or Raman experiments.\\
\begin{figure}
\includegraphics{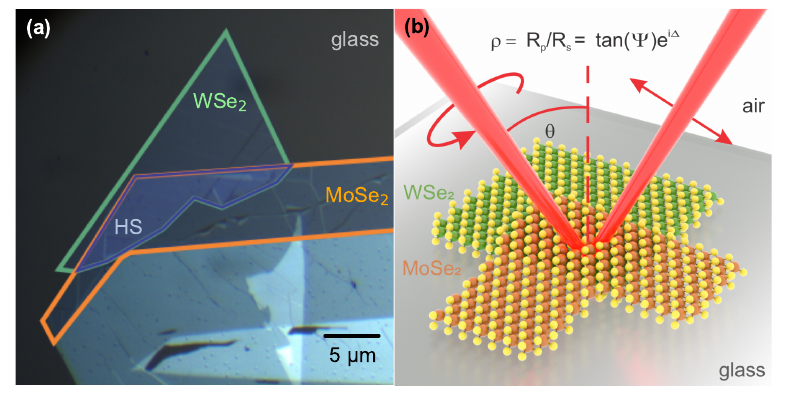}
\caption{\label{Fig1} Spectroscopic imaging ellipsometry (SIE) experiments on MoSe$_2$/WSe$_2$ vdW heterobilayer. (a) Optical microscope image of a MoSe$_2$/WSe$_2$-heterostructure on glass substrate. (b) Schematic description of nulling SIE measurements. The incident elliptical light is adjusted to generate linearly polarized light after being reflected. The analyzer angle is adjusted to satisfy the nulling condition. The polarizer angles satisfy the ellipsometric equation for $\rho$. A resolution of approximately \SI{1}{\micro\meter} allows the investigation of different regions on the sample.}
\end{figure}
An optical micrograph of the investigated MoSe$_{2}$/WSe$_{2}$ heterobilayer is shown in \hyperref[Fig1]{Fig. 1 (a)}. The bulk crystals (hqgraphene) were cleaved onto the PDMS with Nitto white tape. The monolayers were identified via optical contrast and placed on top of each other on a Borofloat 33 (Schott) silicate glass substrate utilizing a viscoelastic stamping method \cite{CastellanosGomez.2014}.\\
SIE measurements are performed in ambient condition using an EP4 (Accurion) ellipsometer with a fixed angle of incidence of $\SI{50}{\degree}$ \cite{Funke.2016}. The ellipsometer is configured in a polarizer (P), compensator (C), sample (S) and analyzer (A), or so-called PCSA geometry. The measurements are performed in nulling mode \cite{Fujiwara.2007}, for which elliptically polarized light is created by the P/C-polarizers such that linearly polarized light is reflected from the surface as sketched in \hyperref[Fig1]{Fig. 1 (b)}. The P- and A-angles are varied with a fixed C-angle to identify the angle configuration resulting in minimal signal on the detector that is either an indivdiual pixel or binned pixels from a charge coupled device (CCD) camera enabling the imaging functionality. The P and A angles for minimal signal determine the ellipsometric angles $\Delta$ and $\Psi$, which satisfy the ellipsometric equation for the complex reflectance matrix
\begin{equation}
    \rho = \frac{R_p}{R_s} = \tan(\Psi)e^{i\Delta},
\end{equation}
with the parallel/orthogonal reflected intensity $R_p$/$R_s$ normalized to the incident light intensity. In the detection arm, a 50x objective with NA=0.45 is used to image the signal onto the CCD camera providing an overall lateral resolution of the setup of $\sim \SI{1}{\micro\meter}$ \cite{Funke.2016}. Tunable and monochromatic illumination is performed by a supercontinuum white light laser combined with acousto-optic tunable filters (AOTF) providing a spectral line-width better than \SI{2}{\nano\meter}. The accessible spectral range covers $\SI{450}{\nano\meter}$ ($\sim \SI{2.76}{\electronvolt}$) to $\SI{1000}{\nano\meter}$ ($\sim \SI{1.24}{\electronvolt}$). Reflections from the backside of the transparent substrate are suppressed with a beam cutter in the excitation path \cite{Funke.2016}. Suitable regions of interest (ROIs) are defined on homogeneous regions of the vdW heterobilayer as well as on individual layers. The ROIs allow the measurement of the ellipsometric spectra $\Psi(\lambda)$ and $\Delta(\lambda)$ with the desired spatial resolution and an optimized signal-to-noise ratio by binning several pixels \cite{Funke.2016}.\\
\begin{figure}
\includegraphics{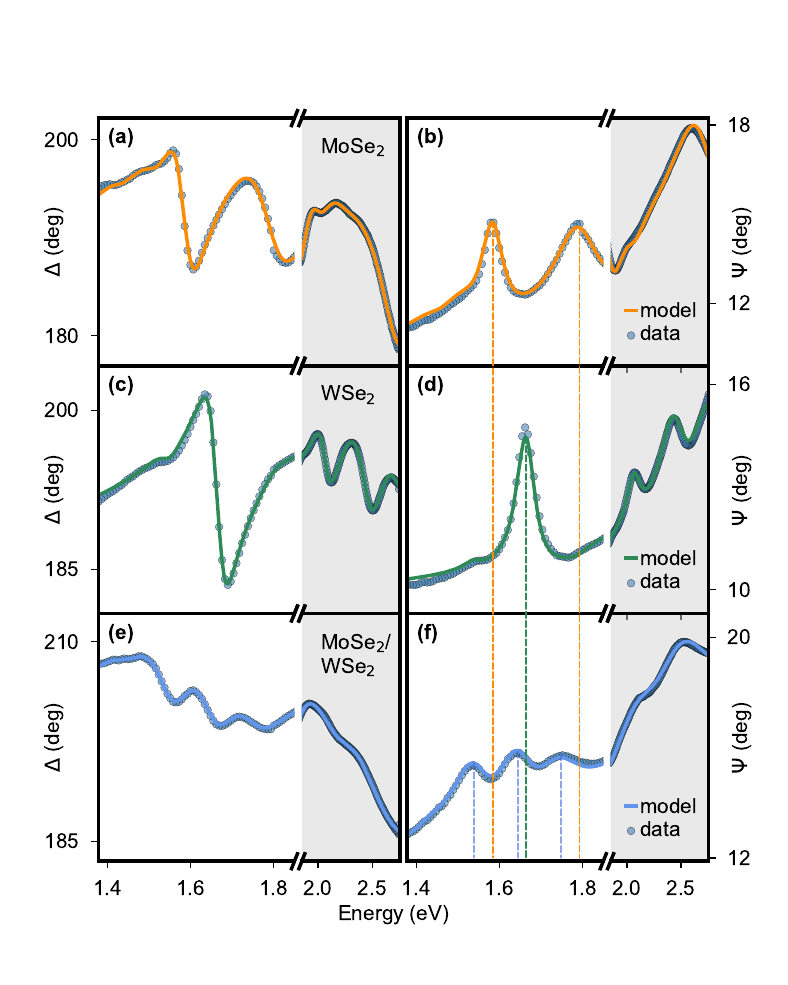}
\caption{\label{Fig2} Comparison of experimentally obtained ellipsometric spectra (grey dots) and fit to the model (solid lines). (a), (c), (e) $\Delta$-values for MoSe$_2$, WSe$_2$ and MoSe$_2$/WSe$_2$ heterobilayer. (b), (d), (f) Respective $\Psi$-values. The spectral range of the A/B excitonic resonances is expanded for better visibility, as indicated by the axes break. In the grey shaded higher energy region the heterobilayer cannot be described as a stack of independent monolayers.}
\end{figure}
To explore the impact of heterostructuring on the light-matter interaction, we determine the complex dielectric functions of the individual MoSe$_2$, WSe$_2$ monolayers and the combined MoSe$_2$/WSe$_2$ heterobilayer by means of SIE. In order to describe the measured SIE spectra of the monolayer and heterobilayer, a comprehensive optical multilayer model is constructed. A fit procedure via regression analysis allows to extract the dielectric function and thickness for each layer \cite{Funke.2016}. The multilayer model consists of a glass substrate, the TMDC layer(s) and air. The glass substrate is approximated as a semi-infinite thickness layer with a Cauchy type dielectric function. The TMDC monolayers on top of the substrate are described by a sum of Tauc-Lorentzian and Lorentzian oscillators to account for the excitonic nature of the resonances \cite{Funke.2016}. \hyperref[Fig2]{Figure 2} provides a comparison of the measured ellipsometric spectra $\Delta(\lambda)$ and $\Psi(\lambda)$ (grey dots) together with the result from regression analysis (solid lines) allowing to extract the dielectric functions for each constituent layer. We find that in order to achieve good agreement between measured ellipsometric spectra and the fit to the model, the MoSe$_2$/WSe$_2$ heterostructure needs to be modelled as an artificial homobilayer with a model that is partially independent of the individual monolayers. By this approach, we achieve very good agreement between measured data and fit to the model as can be seen in \hyperref[Fig2]{Fig. 2}. Minor deviations at the boundaries of the experimentally accessible spectral range can be attributed to out-of-range resonances in the dielectric functions that are included using literature values \cite{Kravets.2017}. The finding that the vdW heterobilayer cannot be properly described by a multilayer system of two independent monolayers can be explained by significant  hybridization effects between the MoSe$_2$ and WSe$_2$ monolayers \cite{Gillen.2018,Kiemle.2020}. This is particularly prominent in the higher energy region (above \SI{1.95}{\electronvolt}) (grey regions in \hyperref[Fig2]{Fig. 2}). The heterobilayer needs to be treated as a new artificial vdW solid rather than an independent stack of TMDC monolayers. A table containing all fit parameters can be found in the supplementary information.\\
\begin{figure}
\includegraphics{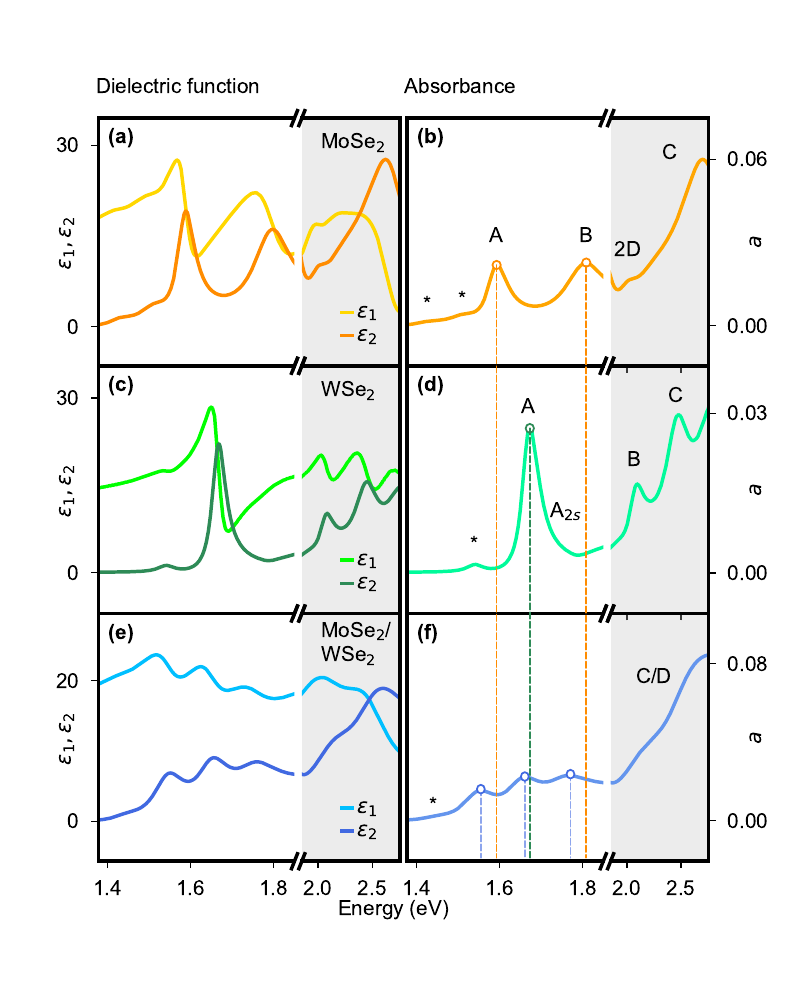}
\caption{\label{Fig3} Dielectric functions and absorbance spectra. (a), (c), (e) Real and imaginary parts of the extracted dielectric functions. The spectral range of the A/B excitonic resonances is expanded for better visibility, as indicated by the axes break. (b), (d), (f) Extracted absorbance spectra. Circles mark the position of the constituent A/B excitonic resonances. A general redshift of the peak positions from monolayer to heterostructure is observed. As higher energetic excitonic resonances can not be directly translated to excitonic resonances in the heterostructure, the heterobilayer is modelled as an artificial homobilayer.}
\end{figure}
In \hyperref[Fig3]{Fig. 3 (a), (c), (e)} the extracted dielectric functions $\epsilon = \epsilon_1 + \epsilon_2$ of MoSe$_2$, WSe$_2$ and the MoSe$_2$/WSe$_2$ heterobilayer are plotted, while \hyperref[Fig3]{Fig. 3 (b), (d), (f)} display the therefrom deduced absorbance spectra $a$ (see supplementary information for details).
The extracted dielectric function of MoSe$_2$ contains two broad and weak sub-gap peaks at \SI{1.426}{\electronvolt} and \SI{1.512}{\electronvolt} marked with asterisks in \hyperref[Fig3]{Fig. 3 (b)} that we attribute to a sub-gap continuum of states related to defect sites \cite{RefaelyAbramson.2018}. Two strong excitonic resonances are assigned to the A- and B-exciton transition at near the fundamental band-gap from the spin-orbit split bands at the K-points of the Brillouin-zone occurring at \SI{1.585}{\electronvolt} and at \SI{1.793}{\electronvolt}, respectively. A weaker resonance marked with a 2D at \SI{1.996}{\electronvolt} is attributed to an interband transition of an unbound electron-hole pair indicating the fundamental band gap \cite{Klein.2019b}. At higher energies the so called C-exciton signatures originating from band nesting between the M and $\Gamma$ points of the Brillouin-zone are described by a sum of several smaller excitonic resonances \cite{Li.2014,Morozov.2015,Hsu.2019}.\\
Analogous to MoSe$_2$, also for WSe$_2$ a weak sub-gap signal related to defects is observed at \SI{1.541}{\electronvolt} (cf. \hyperref[Fig3]{Fig. 3 (d)}). The A-exciton energy is at \SI{1.665}{\electronvolt} and the B-exciton can be seen at \SI{2.067}{\electronvolt}. A signature of the fundamental band gap is expected at approximately \SI{2.1}{\electronvolt} and not resolvable since it is superimposed by the B-exciton \cite{Heienbuttel.2021}. An additional peak between the A- and B-exciton at \SI{1.764}{\electronvolt} can be interpreted as the first excited Rydberg state A$_{2s}$ \cite{Chernikov.2014}. In contrast to MoSe$_2$, the C-exciton is well described by a single peak at \SI{2.496}{\electronvolt}. The fit to the data reveals signatures of the energetically higher-lying D-exciton at approximately \SI{2.768}{\electronvolt} even though this excitonic resonance is not within the directly accessible experimental range.\\
Similar to the constituent monolayers, the sub-gap part of the dielectric function of the MoSe$_2$/WSe$_2$ heterobilayer is best described by a weak, defect related  peak at \SI{1.442}{\electronvolt} (cf. \hyperref[Fig3]{Fig. 3 (f)}). The three distinct peaks in the A/B-exciton spectral range between \SI{1.5}{\electronvolt} and \SI{1.8}{\electronvolt} are attributed to the A-exciton of MoSe$_2$ at \SI{1.543}{\electronvolt}, the A-exciton of WSe$_2$ at \SI{1.647}{\electronvolt} and the B-exciton of MoSe$_2$ at \SI{1.749}{\electronvolt}. The observed redshift of the signatures in comparison to the monolayers are indicated by dashed lines in \hyperref[Fig3]{Fig. 3 (f)}. The A/B-excitons of MoSe$_2$ shift by about \SI{42}{\milli\electronvolt} and \SI{44}{\milli\electronvolt}, respectively. The shift of the A-exciton of WSe$_2$ is \SI{18}{\milli\electronvolt}, while the B-exciton is superimposed by higher lying transitions and no longer clearly resolvable. These observed redshifts are consistent with photoluminescence studies on this material system \cite{Rigosi.2015} and are supposed to be caused by a modified dielectric environment which, combined with layer hybridization, reduces the exciton binding energies \cite{Chernikov.2014,Raja.2017} and stimulates band gap renormalization \cite{Rigosi.2015} due to the transition from monolayer to heterobilayer. As demonstrated above, energetically higher lying transitions lose the single monolayer character and can only be described by a new homogeneous layer with the thickness of the heterobilayer. This result suggests strong hybridization between the two layers, resulting in a delocalization of the electronic states in the heterobilayer, making a description as new artificial vdW solid necessary \cite{Gillen.2018,Kiemle.2020}. The dielectric function in the C/D-exciton spectral range of the individual monolayers is well described by new interband transitions between the hybridized electronic bands of the new vdW solid formed by the MoSe$_{2}$/WSe$_{2}$ heterobilayer. Access to the data sets are provided in the supplementary information as a separate text file.\\
Next, we utilized the obtained dielectric functions to model the Fresnel-based modifications of the reflectance contrast of individual films and MoSe$_2$/WSe$_2$ heterobilayers on commonly used Si/SiO$_2$ substrate materials. The modified interaction with light of the 2D material on a substrate is an important parameter to consider for both the identification by optical contrast as well as optical experiments including reflectance measurements, PL- and Raman-spectroscopy. These modifications are based on the dielectric functions and thicknesses of the investigated multilayer system and therefore have a strong dispersion, making it difficult to e.g. interpret and compare intensities in optical experiments such as the quenching of intralayer excitons which is often considered since it serves as a fingerprint for strong interlayer coupling\cite{Deilmann.2020}.
\begin{figure*}
\includegraphics{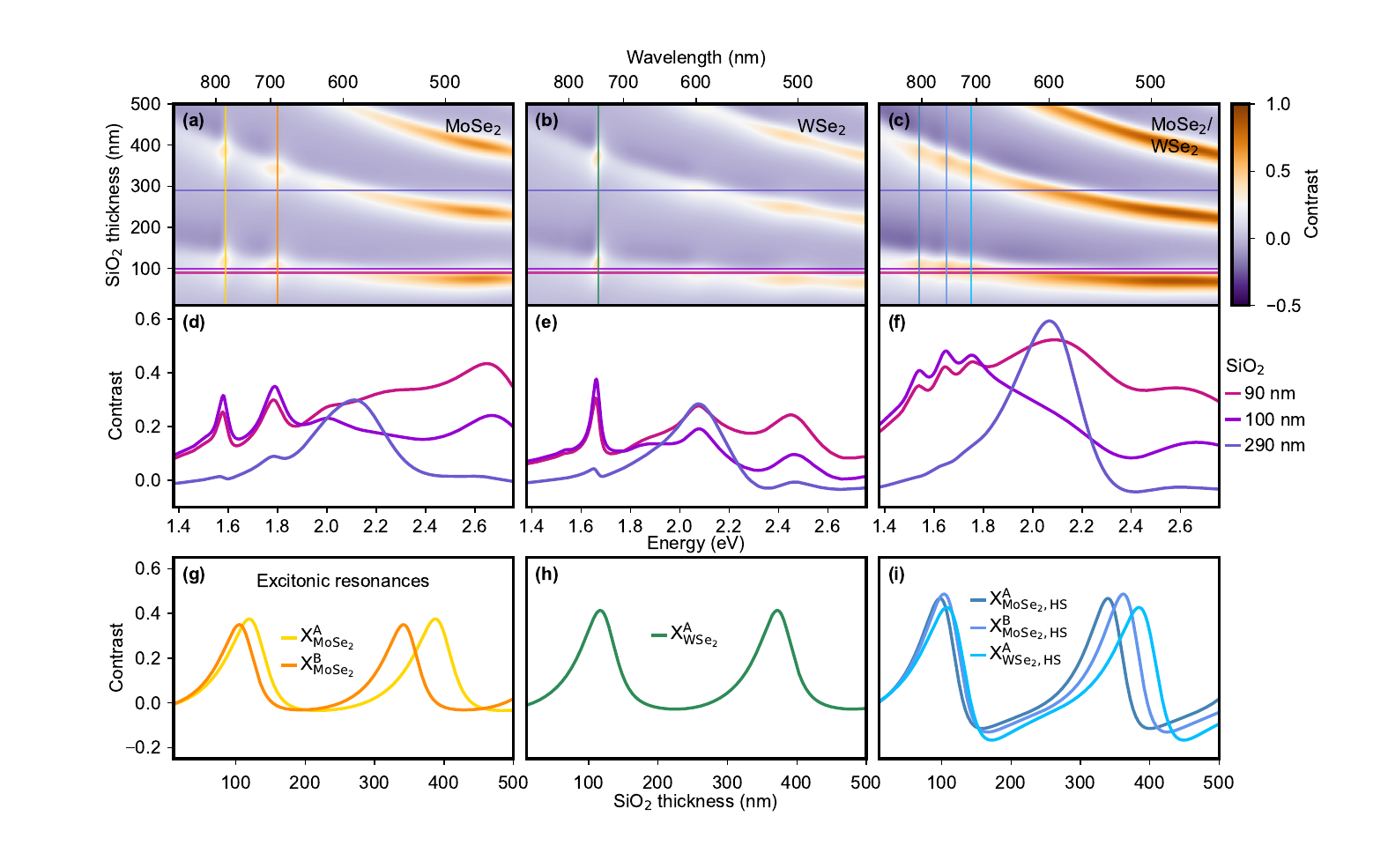}
\caption{\label{Fig4} Simulated Michelson contrast of MoSe$_2$/WSe$_2$ on SiO$_2$ substrates. (a) - (c) Oxide thickness dependent Michelson contrast $C = (R_{subs} - R_{2D})/(R_{subs} +R_{2D})$. Positive values indicate the flake appearing darker compared to the substrate. (d) - (f) Contrast curves for the common \SI{90}{\nano\meter}, \SI{100}{\nano\meter} and \SI{290}{\nano\meter} SiO$_2$-layer substrates. The \SI{290}{\nano\meter} substrate performs significantly worse for MoSe$_2$,WSe$_2$ and their heterostructure. (g) - (i) Contrast curves for the A/B excitonic resonances.}
\end{figure*}
To this end, we employ a plane-wave transfer matrix model to provide the framework of description for the optical response of a multilayer thin film dielectric system \cite{Yeh.1998,Teo.2008}. As base layer, the common Si/SiO$_2$ substrate is used with literature dielectric functions \cite{Malitson.1965,Green.2008}. The Si-layer is treated as semi-infinite. For the TMDCs on top, the dielectric functions obtained from our ellipsometry measurements (cf. \hyperref[Fig3]{Fig. 3}) are used in the calculation. The Michelson contrast can be calculated from the obtained reflectivity by $C = (R_{subs} - R_{2D})/(R_{subs} +R_{2D})$.\\
\hyperref[Fig4]{Figure 4 (a) - (c)} depict the calculated contrast in dependence of the SiO$_2$-thickness between \SI{10}{\nano \meter} and \SI{500}{\nano\meter} and as a function of the experimentally relevant wavelength range from \SI{450}{\nano\meter} to \SI{900}{\nano\meter}. Warmer colors indicate a positive contrast, while colder colors indicate a negative contrast. For all three layer configurations, the excitonic features of the dielectric functions presented in \hyperref[Fig3]{Fig. 3} are clearly visible. The horizontal lines indicate three of the most broadly used SiO$_2$-thicknesses in the 2D community, in particular \SI{90}{\nano\meter},\SI{100}{\nano\meter} and \SI{290}{\nano\meter}. \hyperref[Fig4]{Figure 4 (d) - (f)} presents the contrast curves for those substrates. Both the \SI{90}{\nano\meter} and \SI{100}{\nano\meter} SiO$_2$ layers provide reasonable contrast for MoSe$_2$ and WSe$_2$ as well as their heterostructure, with clearly distinguishable excitonic features. Substrates with a \SI{90}{\nano\meter} thick SiO$_2$ layer have a significantly larger contrast in the sub \SI{700}{\nano\meter} spectral range. The very commonly used \SI{290}{\nano\meter} oxide-layer establishes itself as a suitable substrate for the identification of MoS$_2$ and WS$_2$. However, for the MoSe$_2$/WSe$_2$ heterobilayer investigated in this work, this oxide thickness yields a very poor contrast in the above \SI{700}{\nano\meter} spectral range where interlayer exciton emission is dominant, with the excitonic features having very weak intensities due to destructive interference in the multilayer system in all optical experiments. This result highlights the importance of the choice of the substrate for a specific experiment. The impact of the Fresnel interference in a certain spectral range dominated by the substrate and the whole multilayer structure has to be taken into account for the evaluation of the intensities of the absorbed as well as emitted or scattered light. Since many optical experiments are carried out with detection and excitation in the spectral range of the A and B excitonic resonances of the TMDCs, \hyperref[Fig4]{Fig. 4 (g) - (i)} gives an overview of the substrate dependence on the contrast for the respective excitonic resonances. The spectral cuts are indicated by horizontal lines in \hyperref[Fig4]{Fig. 4 (a) - (c)}. Optimal performance can be achieved by using either an oxide thickness around \SI{100}{nm} or around \SI{300}{nm} to \SI{400}{nm}.\\
Many experiments compare absolute intensities in reflectance, PL- and Raman-spectroscopy using either excitation and/or detection with different light wavelengths. In both cases our simulations demonstrate the sine qua non of taking the choice of substrate into account in the interpretation of the data. Substrate effects can cause variations in the light absorption or emission intensities of much more than an order of magnitude without any intrinsic changes of the light-matter interaction of the investigated materials. With the experimentally determined dielectric functions in particular in the spectral range of the vdW solid that cannot be described by the individual monolayer, our results indicate a guide for better identification and hence fabrication of complex vdW heterostructures. The simulations of the interference contrast can easily be extended to more complex heterostacks including e.g. hBN and graphene for encapsulation and gating \cite{Novoselov.2016,Kiemle.2020}. We demonstrate that it is important to separate intrinsic modification of the light-matter interaction from extrinsic intensity effects due to substrates and multilayer structures that can vary significantly in different spectral ranges e.g. by comparison of intensities of interlayer and intralayer excitons as a signature for interlayer coupling or evaluation of off- and on-resonant excitation in emission or Raman experiments.

\begin{acknowledgments}
The authors acknowledge the financial support by the Deutsche Forschungsgemeinschaft (DFG) via priority program 2244 (2DMP), via cluster of excellence e-Conversion (EXC 2089/1-390776260) and individual projects HO 3324 / 9-2 and WU 637 / 4-2 and 7-1. J.K. acknowledges support by the Alexander von Humboldt foundation.
\end{acknowledgments}

\section*{Data Availability Statement}

\end{document}


\preprint{AIP/123-QED}

\title[]{Supplementary materials for spectroscopic imaging ellipsometry of two-dimensional TMDC heterostructures}

\newcommand{\WSI}{Walter Schottky Institut and Physics Department, Technical University of Munich, Am Coulombwall 4a, 85748 Garching, Germany}
\newcommand{\MCQST}{Munich Center of Quantum Science and Technology (MCQST), Schellingstr. 4, 80799 Munich, Germany}
\newcommand{\ECon}{Exzellenzcluster e-conversion, Technical University of Munich, Lichtenbergstr. 4a, 85748 Garching, Germany}
\newcommand{\WWU}{Institute of Physics, University of Münster, Wilhelm-Klemm-Str. 10, 48149 Münster, Germany}
\newcommand{\SoN}{Center for Soft Nanoscience (SoN), Busso-Peus-Str. 10, 48149 Münster, Germany}
\newcommand{\MIT}{Department of Materials Science and Engineering, Massachusetts Institute of Technology, Cambridge, MA 02139, USA}

\author{Florian Sigger}
    \affiliation{\WSI}
    \address{\MCQST}
    \address{\ECon}
\author{Hendrik Lambers}
    \affiliation{\WWU}
\author{Nisi Katharina}
    \affiliation{\WSI}
    \address{\MCQST}
    \address{\ECon}
\author{Julian Klein}
    \affiliation{\WSI}
    \address{\MIT}
\author{Nihit Saigal}
    \affiliation{\WWU}
\author{Alexander W. Holleitner}
    \affiliation{\WSI}
    \address{\MCQST}
    \address{\ECon}
\author{Ursula Wurstbauer}
    \affiliation{\WWU}
    \address{\SoN}
    \address{\ECon}
    \email{wurstbauer@wwu.de}

\maketitle

\begin{onecolumngrid}

\section*{Absorbance calculation}

The complex refractive indices $\widetilde{n} = n + i\kappa$ are calculated from the dielectric function $\widetilde{\epsilon} = \epsilon_1 + \epsilon_2$ according to

\begin{equation}
\begin{aligned}
\centering
\kappa^2 &= \frac{1}{2}\left(\left(\epsilon_1^2+\epsilon_2^2\right)^{\frac{1}{2}} - \epsilon_1\right)\\
n^2 &= \frac{1}{2}\left(\left(\epsilon_1^2+\epsilon_2^2\right)^{\frac{1}{2}} + \epsilon_1\right).\\
\end{aligned}
\end{equation}

The absorption coefficients/absorbances are determined by

\begin{equation}
\centering
\alpha = \frac{4\pi \kappa}{\lambda}
\end{equation}

and

\begin{equation}
\centering
a = \alpha \times d_{2D},
\end{equation}

where $\kappa$ is the imaginary part of the refractive index, $\alpha$ is the absorption coefficient, $\lambda$ is the wavelength and $d_{2D}$ is the respective thickness of the nanomaterial.

\section*{Fit parameters}

\begin{table}[!h]
\centering
\begin{tabular}{llll|llll|llll}
MoSe$_2$ &        &          &       & WSe$_2$ &         &       &       & \multicolumn{3}{l}{MoSe$_2$/WSe$_2$}   &    \\ 
\hline
E$_0$    & A      & $\Gamma$ & E$_g$ & E$_0$    & A      & $\Gamma$ & E$_g$  & E$_0$    & A      & $\Gamma$ & E$_g$    \\
1.426    & 8.340  & 0.065    & 1.306 & 1.541   & 0.057   & 0.048 & 0.027 & 1.442 & 3.219   & 0.078 & 1.264             \\
1.512    & 0.447  & 0.084    & 0.420 & 1.665   & 118.465 & 0.040 & 1.523 & 1.543 & 16.459  & 0.092 & 1.260             \\
1.585    & 88.282 & 0.051    & 1.426 & 1.764   & 392.647 & 0.132 & 1.778 & 1.647 & 37.608  & 0.096 & 1.439             \\
1.793    & 16.870 & 0.109    & 1.264 & 2.067   & 46.011  & 0.145 & 1.791 & 1.749 & 39.415  & 0.124 & 1.554             \\
1.996    & 11.770 & 0.211    & 1.437 & 2.447   & 5.519   & 0.253 & 0.712 & 2.122 & 49.651  & 0.619 & 1.388             \\
2.228    & 18.583 & 0.443    & 1.358 & 2.767   & 40.017  & 0.375 & 1.879 & 2.508 & 180.711 & 0.644 & 1.997             \\
2.496    & 43.253 & 0.568    & 1.535 & 3.336   & 20.544  & 0.001 & 0.117 & 3.269 & 382.729 & 2.468 & 2.544             \\
2.645    & 26.391 & 0.372    & 1.445 &         &         &       &       &       &         &       &                   \\
3.165    & 30.898 & 0.003    & 1.316 &         &         &       &       &       &         &       &                   \\                    
\end{tabular}
\caption{\label{Tab1} Dielectric function model parameters. All peaks were fitted with a Tauc-Lorentzian line shape. Each column represents the complete model of the respective material.}
\end{table}

\end{onecolumngrid}